\documentclass[12pt,letter,useAMS,natbib]{emulateapj-rtx4}
\usepackage{graphicx,amssymb}
\usepackage{txfonts}
\usepackage{natbib}

\bibpunct[, ]{(}{)}{;}{a}{}{,}
\def \aj {AJ}
\def \mnras {MNRAS}
\def \pasp {PASP}
\def \apj {ApJ}
\def \apjs {ApJS}
\def \apjl {ApJL}
\def \aap {A\&A}
\def \nat {Nature}
\def \araa {ARAA}

\newcommand{\kms} {$\mathrm{ km \; s^{-1}}\,$}

\def\lesssim{\mathrel{\hbox{\rlap{\hbox{\lower4pt\hbox{$\sim$}}}\hbox{$<$}}}}
\def\gtrsim{\mathrel{\hbox{\rlap{\hbox{\lower4pt\hbox{$\sim$}}}\hbox{$>$}}}}

\long\def\symbolfootnote[#1]#2{\begingroup%
\def\thefootnote{\fnsymbol{footnote}}\footnote[#1]{#2}\endgroup} 
\begin{document}


\title{The Unification of Asymmetry Signatures of Type Ia Supernovae}
\author{Justyn R. Maund\altaffilmark{1,2,3}, Peter
  H\"{o}flich\altaffilmark{4}, Ferdinando Patat\altaffilmark{5},
  J. Craig Wheeler\altaffilmark{6}, Paula Zelaya\altaffilmark{7},\\
  Dietrich Baade\altaffilmark{5}, Lifan Wang\altaffilmark{8},
  Alejandro Clocchiatti\altaffilmark{7} and Jason Quinn\altaffilmark{7}} 
\altaffiltext{1}{Dark Cosmology Centre, Niels Bohr Institute, University of Copenhagen, Juliane Maries Vej, DK-2100 Copenhagen \O, Denmark; justyn@dark-cosmology.dk}

\altaffiltext{2}{Department of Astronomy \& Astrophysics, University of California, Santa Cruz, 95064, U.S.A.}

\altaffiltext{3}{Sophie \& Tycho Brahe Fellow}

\altaffiltext{4}{Department of Physics, Florida State University, Tallahassee, Florida 32306-4350, U.S.A.; pah$@$astro.physics.fsu.edu}

\altaffiltext{5}{ESO - European Organisation for Astronomical Research in the Southern Hemisphere, Karl-Schwarzschild-Str.\ 2, 85748 Garching b.\ M\"unchen, Germany; fpatat$@$eso.org; dbaade$@$eso.org}

\altaffiltext{6}{Department of Astronomy and McDonald Observatory, The University of Texas, 1 University Station C1402, Austin, Texas 78712-0259, U.S.A.; wheel$@$astro.as.utexas.edu}

\altaffiltext{7}{Departamento de Astronom$\mathrm{\acute{i}}$a y
  Astrof$\mathrm{\acute{i}}$sica, PUC Casilla 306, Santiago 22, Chile;
  aclocchi$@$astro.puc.cl; pazelaya$@$astro.puc.cl}

\altaffiltext{8}{Department of Physics, Texas A\&M University,
College Station, Texas 77843-4242, U.S.A.; wang$@$physics.tamu.edu}

\begin{abstract}
  We present a compilation of the geometry measures acquired using
  optical and IR spectroscopy and optical spectropolarimetry to probe
  the explosion geometry of Type Ia SNe.  Polarization measurements
  are sensitive to asymmetries in the plane of the sky, whereas line
  profiles in nebular phase spectra are expected to trace asymmetries
  perpendicular to the plane of the sky.  The combination of these two
  measures can overcome their respective projection effects,
  completely probing the 3D structures of these events.  For 9 normal
  Type Ia SNe, we find that the polarization of \ion{Si}{2} $\lambda
  6355$ at 5 days before maximum ($p_{Si\,II}$) is well correlated
  with its velocity evolution ($\dot{\rm v}_{Si\,II}$), implying
  $\dot{\rm v}_{Si\,II}$ is predominantly due to the asymmetry of the
  SNe.  We find only a weak correlation between the
  polarization of \ion{Si}{2} and the reported velocities (${\rm
    v}_{neb}$) for peak emission of optical \ion{Fe}{2} and
  \ion{Ni}{2} lines in nebular spectra.  Our sample is biased, with
  polarization measurements being only available for normal SNe which
  subsequently exhibited positive (i.e. redshifted) ${\rm v}_{neb}$.
  In unison these indicators are consistent with an off-centre delayed
  detonation, in which the outer layers are dominated by a spherical
  oxygen layer, mixed with an asymmetric distribution of intermediate
  mass elements.  The combination of spectroscopic and
  spectropolarimetric indicators suggests a single geometric
  configuration for normal Type Ia SNe, with some of the diversity of
  observed properties  arising from orientation effects.
\end{abstract}

\section{Introduction}
\label{sec:intro}
The importance of Type Ia Supernovae (SNe Ia) as cosmological distance
indicators has been demonstrated over the last twelve years
\citep{1998AJ....116.1009R,1999ApJ...517..565P}.  Key to their
utilisation as probes of the Universe's expansion is an understanding
of their apparent homogeneity (with a magnitude dispersion at maximum
light of $\sigma_{B} \sim 0.8$; \citealt{1993ApJ...413L.105P}) and
the role of the underlying explosion mechanism. Models to describe
these events, as the explosion of a carbon-oxygen Chandrasekhar
mass white dwarf in a single or double degenerate binary system
\citep{branIa}, proceed with burning as deflagrations \citep{2004PhRvL..92u1102G,2006A&A...453..203R},
detonations or a combination of the two:
delayed detonations \citep{1991A&A...245..114K}.  These
different mechanisms will imprint a signature on the resulting
chemical abundances and geometries of the ejecta
\citep{2006NewAR..50..470H}.

The geometries of  SNe Ia have been probed using different
techniques.  At early times, spectropolarimetry of these events has
shown a wide array of asymmetries \citep[see][for a
review]{2008ARA&A..46..433W}, in the plane of the sky.  In general, the
continuum polarization, which probes the shape of the photosphere, is
of order a few tenths of a percent, indicating departures from
a spherical symmetry of $<10\%$ \citep{1991A&A...246..481H,1997ApJ...476L..27W}.    Significant line polarization is
observed, showing the distribution of elements in the ejecta, predominantly for \ion{Ca}{2} \citep[e.g. for SN
2001el][]{2003ApJ...591.1110W,2003ApJ...593..788K} and the
principal classification feature for SNe Ia, \ion{Si}{2} $\lambda
6355$.  \citet{2007Sci...315..212W} identified a correlation between
the degree of polarization of \ion{Si}{2} and the light curve decline rate
parameter $\mathrm{\Delta m_{15}(B)}$.

Nebular phase spectroscopy of SNe~Ia in the optical and infrared (IR),
when the ejecta are optically thin and the spectra are dominated
by the Fe-group elements, can be used to reveal the structure at the
centre of the explosion; albeit projected onto the radial velocity
direction
\citep{2006NewAR..50..470H,2006ApJ...652L.101M,2007ApJ...661..995G,2010ApJ...708.1703M}. 
The almost-unblended Fe line at 1.6 $\mu m$ shows
 peculiar line profiles such as flat topped profiles indicating
 material with central cavities, and 
\citet{2007ApJ...661..995G} observed identical profiles and features in mid-IR lines. 
 This strongly supports that the line asymmetries are kinematic in nature and all but exclude
explanations due to uncertainties in the atomic physics, which may
lead to an underestimation of blends,
or optical thickness effects.

\citet{maedapap}  linked the asymmetries of the central layers with
expansion velocities less than $\lesssim 3000$ \kms
with the outer layers.  They found that the peak of the nebular emission line
profiles, in the optical, exhibited blue or red shifts, consistent with an
asymmetric distribution along the line of sight.   It was found that this
kinematical offset was related to the evolution of the velocity of
the absorption minimum of the \ion{Si}{2} $\lambda 6355$ feature
(parametrized by the temporal velocity gradient
$\dot{\rm v}_{Si\,II}$) at early times; suggesting both are signatures of
the departure of SNe~Ia from simple spherical symmetry.

Probing the inner structure and correlations with outer layers is
central to answering key questions of supernova theory \citep[for a
review see][]{2006NuPhA.777..579H}, such as whether the asymmetries in
the inner and outer layers have a common physical origin.  In this
letter we discuss the correlations between geometry indicators derived
from early and late spectroscopic observations and spectropolarimetric
observations of SNe Ia, to approach a unified model of the behaviour
of these events.

\section{Polarization and the Velocity Gradient of \ion{Si}{2}}
\label{sec:siii}
\citet{2007Sci...315..212W} observed that the polarization of the
\ion{Si}{2} line for normal SNe Ia peaks at $\sim -5$ days (relative
to B-light curve maximum), and this was further confirmed for SN~2006X
\citep{2009A&A...508..229P}.  Values of $\dot{\rm v}_{Si\,II}$ for a
range of SNe~Ia were compiled from those reported by
\citet{2005ApJ...623.1011B} and \citet{maedapap}.  These were cross
checked against the list of SNe Ia with the polarization of
\ion{Si}{2} $\lambda 6355$ reported at $-5$ days
\citep{2007Sci...315..212W}.

For the normal SN 2007le, the \ion{Si}{2} polarization was estimated
from spectropolarimetric observations at -9 and -4 days (Zelaya et
al., in prep.), with $\dot{\rm
  v}_{Si\,II}=83\pm3$ \kms$\mathrm{day^{-1}}$
\citep{2009ApJ...702.1157S}.  The polarization of \ion{Si}{2} $\lambda
6355$ for SNe 2002bf and 2009dc, observed at $+3$ days and $+5.6$ days
after B-maximum were acquired from \citet{2005ApJ...632..450L} and
\citet{2010ApJ...714.1209T}, respectively (with values of $\dot{\rm
  v}_{Si II}$ given by \citealt{2005ApJ...632..450L} and
\citealt{2009ApJ...707L.118Y}).  Given that the observations of
SN~2002bf and 2009dc were conducted later than $-5$ days, we assume
that the reported polarizations represent {\it lower limits} of
$p_{Si\,II}$ at $-5$ days.  For SN~2001V we estimated $\dot{\rm
  v}_{Si\,II}=81\pm7$ \kms$\mathrm{day^{-1}}$ \citep[using the data
of][]{2008AJ....135.1598M}\footnote{The data were acquired from the
  SUSPECT archive: http://suspect.nhn.ou.edu/$\sim$suspect/} , which
is higher than the $~20$ \kms$\mathrm{day^{-1}}$ measured for the
similar SN~1991T \citep{1992AJ....103.1632P, 2008MNRAS.385...75T}.

The \ion{Si}{2} polarizations for the peculiarly faint 1991bg-like SNe
1999by and 2005ke were taken from \citet{2001ApJ...556..302H} and
Patat et al., (in prep.), respectively.  The velocity evolution of \ion{Si}{2} in this particular
sub-class has been found to be relatively homogeneous
\citep{2008MNRAS.385...75T}.  We derive an average value $\dot{\rm
  v}_{Si\,II}=105.8\pm8.6$ \kms\,$\mathrm{day^{-1}}$ from the reported
velocities of 4 members of the subluminous subclass (SNe 1991bg
\citealt{1996MNRAS.283....1T}; 1999by
\citealt{2001AJ....121.3127V,2001ApJ...556..302H}; 2005bl
\citealt{2008MNRAS.385...75T}; and 2005ke Patat et al., in prep.).  In
the absence of further spectroscopic data, we assume that this value
of the deceleration of \ion{Si}{2} is valid for SN~2005ke.

\begin{figure}
\includegraphics[width=8cm]{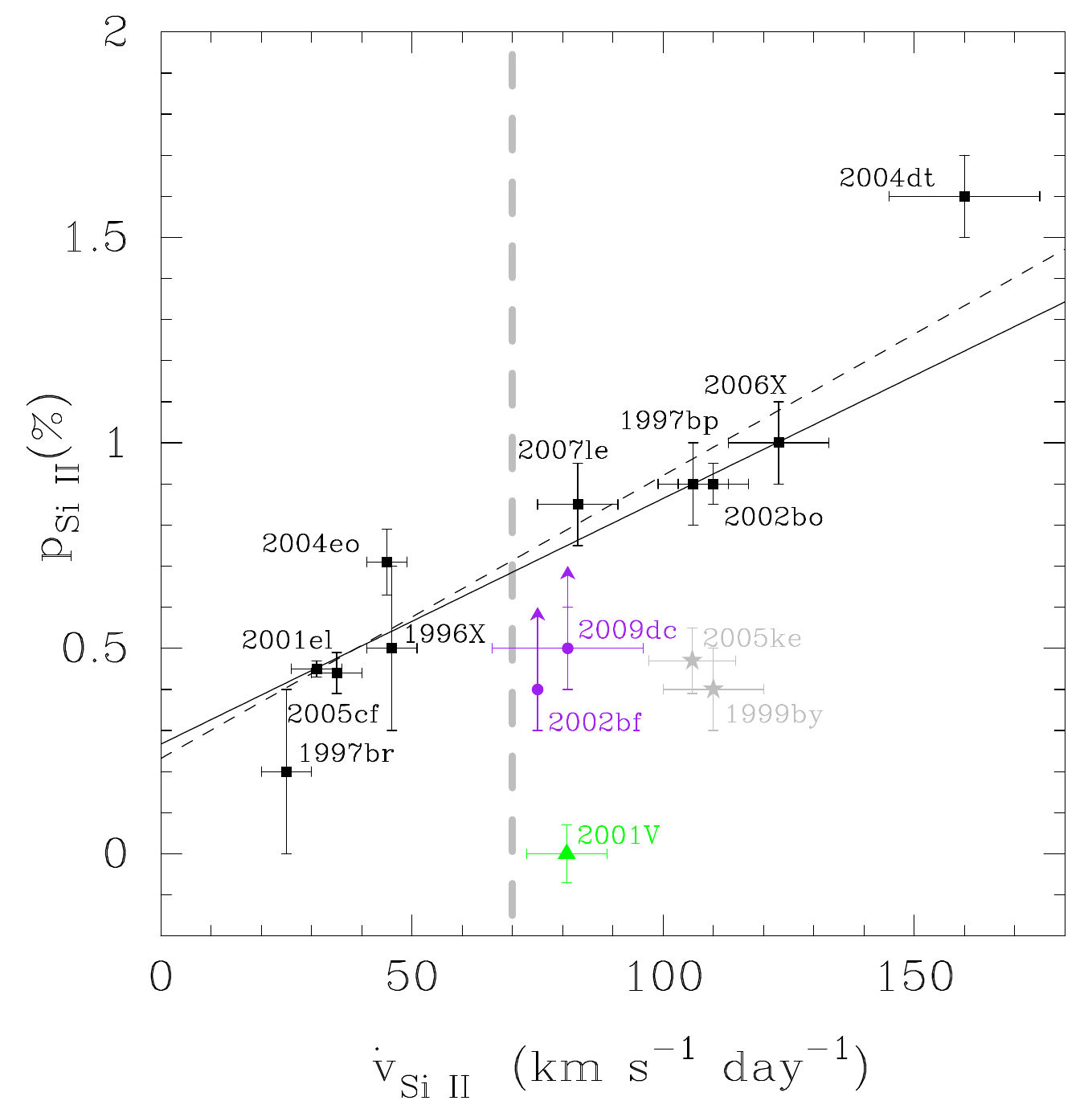}
\caption{The observed line polarization of \ion{Si}{2} $\lambda
  6355$ at $-5$ days, relative to B light-curve maximum, as a function
  of the measured velocity gradient for
 the same line.  The group of normal SNe Ia are indicated in black ($\blacksquare$), whereas the peculiarly faint SNe 1999by and SN 2005ke are
  indicated in grey ($\star$) and the single representative
  of the SN~1991T-like SNe 2001V is indicated in green ($\blacktriangle$).  Lower limits on the polarization of
  \ion{Si}{2} for two SNe are presented in purple ($\bullet$).   The dashed grey
  line at 70 \kms$\mathrm{day^{-1}}$ separates the HVG and LVG SNe~Ia
  \citep{2005ApJ...623.1011B}.  The solid black line indicates the
  best-fit straight line to the normal SNe Ia, while the dashed black
  line is for the best-fit including SN~2004dt.}
\label{fig:specpol:vdot}
\end{figure}

The relation between the velocity gradient of \ion{Si}{2} and the
associated polarization is shown on Fig. \ref{fig:specpol:vdot}.
There is a distinct correlation for normal SNe~Ia and, excluding SN~2004dt, we find $p_{Si II}
= 0.267 + 0.006 \times \dot{\rm v}_{Si II}$ (with
$\chi^{2}_{\nu}=1.07$, Pearson correlation coefficient $r=0.93$).  The
SN~1991T-like SN~2001V and the faint SNe 1999by and 2005ke are
obvious outliers from this relation.  In comparing $p_{Si\,II}$ and
the light-curve decline rate parameter $\Delta m_{15}(B)$,
\citet{2007Sci...315..212W} found that these SNe are significant
outliers from the linear correlation found between these two
parameters.  SN~2004dt is not a significant outlier ($<3\sigma$), and
including it in the fit slightly changes the form of the
relation.  It may indicate that the relation at very high $\dot{\rm
  v}_{Si\,II}$ is no longer linear.

The dividing line between the High Velocity Gradient (HVG) and Low
Velocity Gradient (LVG) SNe is a deceleration of $\sim 70$ \kms
$\mathrm{day^{-1}}$ in Fig. \ref{fig:specpol:vdot}
\citep[see][]{2005ApJ...623.1011B}.  The relative numbers in each group,
however, are dictated by the selection effect of the number of SNe
with both sufficiently dense early spectroscopy, to derive the
\ion{Si}{2} velocity evolution, and early spectropolarimetry.  The gap
between the HVG and LVG SNe may not, therefore, be representative of
significant differentiation between the two types of normal SNe~Ia.
The lower polarization limits provided by SNe 2002bf and 2009dc may
fill this gap and indicate that SNe Ia form a continuous distribution
on this diagram.  Caution is required, however, as
\citet{2005ApJ...632..450L} suggest SN~2002bf may be related to the
peculiar SN~2004dt, due to its high velocities, and
\citet{2009ApJ...707L.118Y} and \citet{2010ApJ...714.1209T} suggest
SN~2009dc may have arisen from a super-Chandrasekhar mass White Dwarf
progenitor.

\section{Polarization and the Nebular Phase Velocity of Fe-group Elements}
\label{sec:vneb}
We examined the compilation of line-of-sight velocity offsets ${\rm
  v}_{neb}$ for the peaks of the emission lines of $[$\ion{Fe}{2}$]$
$\lambda 7155$ and $[$\ion{Ni}{2}$]$ $\lambda 7378$, relative to their
rest wavelength, provided by \citet{maedapap}.  During the nebular
phase ($t > 100$ days) these are presumed to be a measure of the
expansion velocity of the central portions of the ejecta,
representative of the products of the deflagration phase.  For
SN~2005df, we utilised the measurements of \citet{2007ApJ...661..995G}
of the velocity offsets of $[$\ion{Ni}{3}$]$ 7.35 and
11.002$\mathrm{\mu m}$\ and $[$\ion{Ni}{4}$]$ 8.41$\mu m$\ at 118 days
(although \citeauthor{2007ApJ...661..995G} do not observe an offset
associated with $[$\ion{Co}{3}$]$ 11.89$\mu m$).

The velocity offsets for SN~2001V were determined for the optical
\ion{Fe}{2} and \ion{Ni}{2} lines, from a spectrum acquired at 106
days \citep{2008AJ....135.1598M}.  We measured, however, different
velocity offsets for the two species.  This may indicate a more
complicated differentiation of the distributions of Fe and Ni than
observed for the normal HVG and LVG SNe Ia; and may be commensurate
with SN~2001V belonging to the subclass of 1991T-like SNe~Ia.

\begin{figure}
\includegraphics[width=8cm]{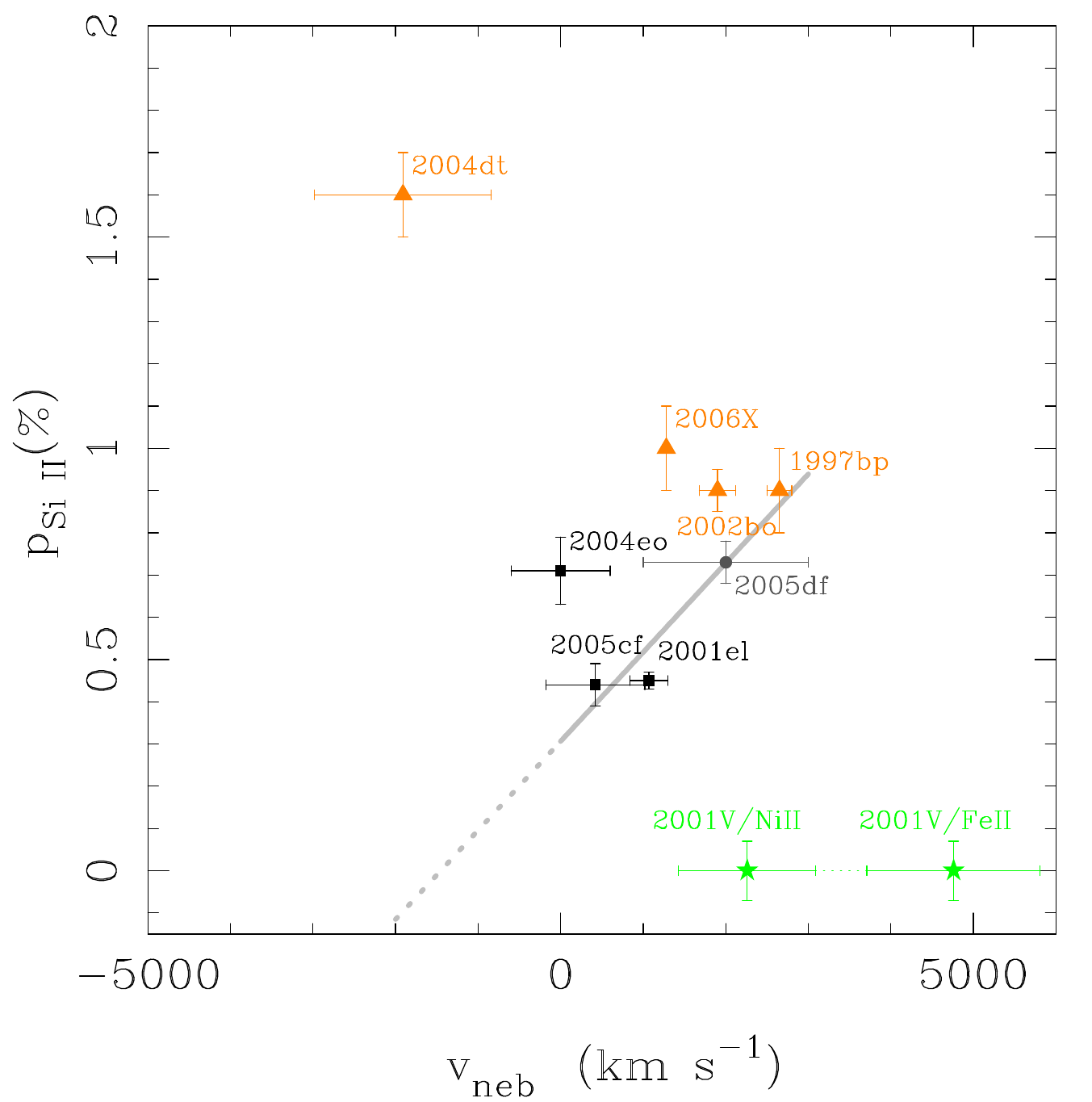}
\caption{The observed line polarization of  \ion{Si}{2} $\lambda
  6355$ at $-5$ days as a function of the nebular phase velocity offset.  HVG SNe are
  indicated in orange ($\blacktriangle$), LVG SNe are in black
  ($\blacksquare$), SN~2005df (without a measurement of
  $\dot{\rm v}_{Si\,II}$) is shown in grey ($\bullet$) and the two
  measurements for SN~2001V are shown in green ($\star$).  An
  approximate trend between $p_{Si\,II}$ and ${\rm v}_{neb}$ for normal
      SNe~Ia is indicated by the
  grey line.}
\label{fig:specpol:neb}
\end{figure}

In Fig. \ref{fig:specpol:neb} the
polarization of \ion{Si}{2} $\lambda 6355$ at early times is compared
with the nebular velocity offset for events for which both
data exist.  There is an obvious sample
bias, however, as the only SN~Ia with a suitable polarization
measurement and a blue shifted emission line profile is SN~2004dt.
The remaining SNe~Ia from our sample all
have redshifted emission lines in their optical nebular spectra.

The apparent separation between the normal HVG and LVG SNe Ia in
Fig. \ref{fig:specpol:neb} approximately reflects the correlation
between $\dot{\rm v}_{Si\,II}$ and $p_{Si\,II}$ established in
\S\ref{sec:siii}, and the observation of \citet{maedapap} that the
nebular lines of SNe Ia with higher values of $\dot{\rm v}_{Si\,II}$
exhibit larger red displacements.  The correlation between ${\rm
  v}_{neb}$ and $p_{Si\,II}$ for normal SNe~Ia, excluding SN~2004dt
and 2001V, is much weaker ($r=0.54$) than the correlation of
Fig. \ref{fig:specpol:vdot}.  In this sample, based on measurements of
${\rm v}_{neb}$ alone, it is difficult to distinguish between LVG and
HVG normal SNe~Ia (e.g. SNe 2001el and 2006X).  Without further SNe
with blue-shifted nebular velocities and polarization measurements, it
is difficult to ascertain the behaviour in the blue-shifted portion of
the diagram.  Given the relationship established in \S\ref{sec:siii},
those LVG SNe with negative ${\rm v}_{neb}$ should appear in the lower
left-hand quadrant of Fig. \ref{fig:specpol:neb}, tending towards zero
polarization.

\citet{2009A&A...505..265L} measured a velocity gradient of
$\dot{\rm v}_{Si\,II}=41\pm6$ \kms$\mathrm{day^{-1}}$ for the normal
SN~2003hv.  This would suggest $p_{Si\,II}\sim 0.5\%$ at -5 days,
which is consistent with adopting the observed polarization at $+6$
days (Maund et al., in prep.) of $0.25\%\pm0.05\%$ as a lower limit.
The implied decrease in the $p_{Si\,II}$ between $-5$ and $+6$
days, however, is much slower than found for other normal SNe~Ia
\citep{2007Sci...315..212W}.  \citet{2006ApJ...652L.101M} and
\citet{2009A&A...505..265L} observed the $[$\ion{Fe}{2}$]$ line
profiles for SN~2003hv as blue-shifted ($-2750\pm440$\kms ;
\citealt{maedapap}).  Depending on the strength of \ion{Si}{2}
polarization at -5 days, if the polarization decreased dramatically,
SN~2003hv may occupy a similar locus to that of SN~2004dt on Fig. \ref{fig:specpol:neb}.  This would
support a trend relating $p_{Si\,II}$ and ${\rm v}_{neb}$ orthogonal
to that apparent from the central group of points on Fig. \ref{fig:specpol:neb}, with SN~2004dt and SN~2001V at the extremes of a
general trend that runs through the cloud of normal HVG and LVG SNe~Ia.  If the
polarization of SN~2003hv did not decrease substantially from -5 to +6
days, then SN~2003hv appears to conform with the trend identified for
normal SNe~Ia.
\section{Discussion \& Conclusions}
\label{sec:disc}
\citet{maedapap} observed a relationship between $\dot{\rm
  v}_{Si\,II}$ and ${\rm v}_{neb}$, and interpreted it as indicating a
single asymmetric geometry for Type Ia~SNe.  This asymmetry gives rise
to the apparent diversity amongst this SN class due to the different
orientations at which individual SNe are observed. In also considering
polarimetric observations, we find a good linear correlation between
$p_{Si\,II}$, an established indicator of geometry, and $\dot{\rm
  v}_{Si\,II}$, implying the latter is a signature of the geometry.
The observed tight correlation between these two parameters for normal
SNe~Ia implies that the asymmetries probed by $p_{Si\,II}$ are
unlikely to be due to a random, clumpy line forming region, rather it
indicates the role of a large scale asymmetry in the ejecta.  We find
that $p_{Si\,II}$ shows a weak correlation with the later nebular
velocity, that may indicate a possible correlation between the
asymmetries inferred for the layers observed at early and late times.
This correlation is expected to be weak, however, as $p_{Si\,II}$ and
${\rm v}_{neb}$ probe orthogonal projections of the geometry.  This
suggests that in trying to understand the influences of geometry on
measurable photometric and spectroscopic parameters, such as $\Delta
m_{15}(B)$, it is preferable to use the early time indicators
$\dot{\rm v}_{Si\,II}$ and $p_{Si\,II}$.

\citet{2005ApJ...623.1011B} suggested the difference between HVG and
LVG SNe~Ia arises from the orientation at which the ejecta are
viewed. They hypothesised that SNe~Ia will be observed as HVG SNe if
the ejecta approaching the observer were mixed with heavy elements,
increasing the opacity and keeping the photosphere at high velocities
at early times \citep{1993A&A...268..570H}.  In their model, LVG
SNe~Ia are observed from other angles, dominated by intermediate mass
elements (IMEs), through which the photosphere has already receded even
at early times.  A key consequence of this model is asymmetric
excitation of the ejecta leading to an asymmetric photosphere, in
particular for HVG SNe.  This is in contrast to the low limits placed
on the departure of the photospheres of SNe~Ia from spherical symmetry
from polarimetry of these events ($p_{cont} \sim 0.2-0.3\%$;
\citealt{2003ApJ...591.1110W,2009A&A...508..229P,1997ApJ...476L..27W}).

It is significant that \ion{Si}{2}, amongst other line features, is
generally observed to be polarized in SNe~Ia, whereas the continuum is
not.  This requires that the line forming region be asymmetrically
distributed across a spherical photosphere
\citep{2003ApJ...593..788K,2005ApJ...632..450L,my2005hk}.\\
\citet{2005ApJ...623.1011B} and \citet{maedapap} interpret the
velocity of \ion{Si}{2} $\lambda 6355$ as the photospheric velocity,
which depends on the degree of mixing of Fe-group elements and the
opacity in the outer layers of the ejecta.
\citet{1996MNRAS.278..111P} find, however, that the \ion{S}{2} ``W''
feature at $5640\AA$, which shows lower velocities and a less severe
velocity gradient than \ion{Si}{2}, more accurately reflects the true
photospheric velocity.  \citet{2006MNRAS.370..299H} observed that the
velocity determined for \ion{S}{2} is lower than that determined from
\ion{Si}{2} $\lambda 6355$ for all SNe types, and that the difference
is largest for the HVG SNe.  This implies \ion{Si}{2} $\lambda 6355$
is formed above the photosphere, and is not specifically coupled with
the photospheric velocity.  The velocity evolution of \ion{Si}{2} is dependent
on the mass contained in the \ion{Si}{2} line forming region above the
photosphere (deposited by the protrusions of IMEs into the outer
layers).  Conversely to the correlation of
\citet{2005ApJ...623.1011B}, an increase in opacity due to Fe-group
elements mixed into the outer ejecta layers would slow the apparent
recession of the photosphere leading to an LVG rather than HVG SN.

For SN~2004dt,  \citet{2006ApJ...653..490W} and
\citet{2005ApJ...632..450L} observed negligible polarization
associated with the \ion{O}{1} $\lambda7774$, which
\citeauthor{2006ApJ...653..490W} interpreted as a spherical
distribution of unburned oxygen evenly covering the photodisk.  That
both \ion{Si}{2} and \ion{O}{1} occupied the same velocity space, but had
very different polarization properties, suggested that portions of
the oxygen layer were mixed with IMEs (such as Si) with an asymmetric distribution
\citep{2006NewAR..50..470H}.  The combination of low polarization for
oxygen and the continuum, but with polarized silicon has been observed
for other SNe~Ia of both LVG and HVG classes such as SNe 2001el, 2006X
and 2009dc
\citep{2003ApJ...591.1110W,2009A&A...508..229P,2010ApJ...714.1209T}.
The uniformity of oxygen in the outer ejecta of SNe Ia is further
demonstrated by \citet{2006MNRAS.370..299H}, who find almost constant
line strength for the absorption of \ion{O}{1} $\lambda 7774$ for both
LVG and HVG SNe at B light-curve maximum.

Based on their correlation, \citet{maedapap} suggest that
instabilities of both the central and outer layers are caused during
the deflagration phase and due to rising plumes.  The geometry, and in
particular the physics of the runaway and the location of the
deflagration to detonation transition, may be affected by other
processes such as the progenitor's rotation and its binary companion
\citep{2001ApJ...556..302H,2006NewAR..50..470H}. In the model of
\citet{maedapap}, significant heavy elements from the deflagration are
mixed into the outer layers, on the side of the ejecta from which an
HVG SN~Ia is observed.  

The polarization of \ion{Si}{2}, and the inferred asymmetries, and the
correlation with $\dot{\rm v}_{Si\,II}$ suggest a single offset, with
HVG SNe~Ia being those with an offset Si distribution mixed into the
outer O layer in the
direction of the observer and the products of deflagration receding.  At significant angles away from the offset
direction, the Si is found in a thinner layer, more evenly excited by
the underlying Ni substrate, leading to an LVG SN~Ia with a lower
polarization for \ion{Si}{2} $\lambda 6355$ and in which products of
deflagration appear to approach the observer.  These correlations are
roughly consistent with the distributions in
Fig. \ref{fig:specpol:vdot} and \ref{fig:specpol:neb}.  The low
continuum polarization for HVG and LVG SNe implies the products of the
detonation, such as $\mathrm{^{56}Ni}$ are spherically distributed in the ejecta.

The two subluminous SNe, 1999by and 2005ke, show low line polarization. Within the
framework of single degenerate scenarios, however, such SNe~Ia produce
${\rm ^{56}Ni}$ in the deflagration, whereas during the detonation phase Si-group elements
are produced at the expense of ${\rm ^{56}Ni}$ \citep{1995ApJ...444..831H,2002ApJ...568..791H}.  In
models of normal SNe~Ia, the majority of the ${\rm ^{56}Ni}$ is produced during the
detonation phase. The fact that line polarization is observed to be strongest
in normal SNe~Ia and small in subluminous SNe~Ia argues against
deflagration instabilities as the origin of the observed correlations.

As polarization probes asymmetries in the plane of the sky and nebular
phase velocities probe asymmetries perpendicular to the plane of the
sky, the combination of both these measures provides the opportunity
to overcome projection effects and completely probe the
three-dimensional structure of these events.  By bringing together the
measurements of $p_{Si\,II}$, $\dot{\rm v}_{Si\,II}$ and ${\rm
  v}_{neb}$, a portion of the spectral
diversity of normal SNe~Ia may be understood in terms of  simple
orientation effects concerning a single geometry and explosion by an off centre
delayed detonation.

\section*{Acknowledgements}
The research of JRM is funded through the Sophie \& Tycho Brahe
Fellowship.  The Dark Cosmology Centre is supported by the DNRF.  The
research of JCW is supported in part by NSF grant AST-0707769. AC, JQ,
and PZ thank the support of Basal CATA PFB 06/09, FONDAP No. 15010003,
and P06-045-F (ICM/MIDEPLAN/Chile).  The authors are grateful to
Stefan Taubenberger for useful discussions concerning the
velocity evolution of the faint subclass of SNe~Ia.
\bibliographystyle{apj}

\end{document}